\documentstyle[
axodraw,
psfig,
manyeqns,
12pt]{article}
\newcommand{\MML}[2]{\hat{M}^{#1}_L (\hat{M}^{#2}_L)^*}
\newcommand{\MMR}[2]{\hat{M}^{#1}_R (\hat{M}^{#2}_R)^*}
\newcommand{\MMaL}[1]{|\hat{M}^{#1}_L|^2}
\newcommand{\MMaR}[1]{|\hat{M}^{#1}_R|^2}

\newcommand{\hdick}{\noalign{\hrule height1.4pt}}

\newcommand{\efft}{{\epsilon_\tau}}
\newcommand{\effb}{{\epsilon_b}}
\newcommand{\h}{{H}} 
\newcommand{\hi}{{\sigma}} 
\newcommand{\hf}{{\sigma'}} 
\renewcommand{\Im}{{\mbox{Im}\,}}
\renewcommand{\Re}{{\mbox{Re}\,}}
\newcommand{\pv}{{q}} 

\renewcommand{\theequation}{\thesection.\arabic{equation}}
\newcommand{\cleqn}{\setcounter{equation}{0}}

\newcommand\commentout[1]{}
\newcommand{\pr}{\hspace{\parindent}}

\def\beq{\begin{equation}}
\def\eeq{\end{equation}}
\def\bea{\begin{eqnarray}}
\def\eea{\end{eqnarray}}
\def\bsub{\begin{subequations}}
\def\esub{\end{subequations}}

\def\simgt{\,{\rlap{\lower 3.5pt\hbox{$\mathchar\sim$}}\raise 1pt\hbox{$>$}}\,}
\def\simlt{\,{\rlap{\lower 3.5pt\hbox{$\mathchar\sim$}}\raise 1pt\hbox{$<$}}\,}

\def\to{\rightarrow}

\def\epem{\ifmmode{ e^{+}e^-} \else{$ e^{+}e^- $ } \fi}
\def\ttbar{\ifmmode{t\bar{t}} \else{$t\bar{t}$} \fi}
\def\ffbar{\ifmmode{f\bar{f}} \else{$f\bar{f}$} \fi}
\def\ppbar{\ifmmode{p\bar{p}} \else{$p\bar{p}$} \fi}
\def\msbar{\ifmmode{\overline{\rm MS}} \else{$\overline{\rm MS}$} \fi}

\def\fb{{\rm FB}}

\def\mz{m_Z^{}}

\def\mh{m_H^{}}

\makeatletter

\newtoks\@stequation

\def\subequations{\refstepcounter{equation}%
  \edef\@savedequation{\the\c@equation}%
  \@stequation=\expandafter{\theequation}
  \edef\@savedtheequation{\the\@stequation}
  \edef\oldtheequation{\theequation}%
  \setcounter{equation}{0}%
  \def\theequation{\oldtheequation\alph{equation}}}

\def\endsubequations{%
  \ifnum\c@equation < 2 \@warning{Only \the\c@equation\space subequation
    used in equation \@savedequation}\fi
  \setcounter{equation}{\@savedequation}%
  \@stequation=\expandafter{\@savedtheequation}%
  \edef\theequation{\the\@stequation}%
  \global\@ignoretrue}

\def\eqnarray{\stepcounter{equation}\let\@currentlabel\theequation
\global\@eqnswtrue\m@th
\global\@eqcnt\z@\tabskip\@centering\let\\\@eqncr
$$\halign to\displaywidth\bgroup\@eqnsel\hskip\@centering
     $\displaystyle\tabskip\z@{##}$&\global\@eqcnt\@ne
      \hfil$\;{##}\;$\hfil
     &\global\@eqcnt\tw@ $\displaystyle\tabskip\z@{##}$\hfil
   \tabskip\@centering&\llap{##}\tabskip\z@\cr}

\makeatother

\newcommand{\eps}{\epsilon}

\newcommand{\bmeq}{\begin{manyeqns}}
\newcommand{\emeq}{\end{manyeqns}}
\newcommand{\mmz}{m_Z^2}
\newcommand{\mmh}{m_\h^2}

\newcommand{\gfL}{{g^{f}_L}}
\newcommand{\gfR}{{g^{f}_R}}

\newcommand{\geL}{{g^{e}_L}}
\newcommand{\geR}{{g^{e}_R}}
\newcommand{\ges}{{g^{e}_\hi}}
\newcommand{\gZ}{{g_{Z}^{}}}
\newcommand{\one}{{\hphantom{.}1\hphantom{00}}}
\newcommand{\zero}{{\hphantom{.}0\hphantom{00}}}
\renewcommand{\fb}{{\mbox{fb}}}

\begin{document}

\begin{flushright}
KEK-TH-663 \\
HUE-99-2 \\
OCHA-PP-146 \\
DESY 99-190 \\
hep-ph/0002043 \\ 
\end{flushright}

\begin{center}
\Large
Prospects of Measuring General Higgs Couplings \\ at $e^+e^-$ Linear Colliders
\vspace{0.5cm}

\large
  K. Hagiwara$^a$,
  S. Ishihara$^{a,b}$,
  J. Kamoshita$^c$, 
  B.A. Kniehl$^d$
\vspace{0.5cm}

\small 
~$^a$  Theory Group, KEK, \\
1-1 Oho, Tsukuba, Ibaraki 305-0801, Japan
\vspace{0.5cm}

~$^b$  Department of Physics, Hyogo University of Education, \\ 
941-1 Shimokume, Yashiro, Kato, Hyogo 673-1494, Japan
\vspace{0.5cm}

~$^c$  Department of Physics, Ochanomizu University, \\
2-1-1 Otsuka, Bunkyo, Tokyo 112-8610, Japan 
\vspace{0.5cm}

~$^d$  II. Institut f\"ur Theoretische Physik, Universit\"at Hamburg, \\
Luruper Chaussee 149, 22761 Hamburg, Germany
\end{center}

\begin{abstract}
We examine how accurately the general $HZV$ couplings, with $V=Z,\gamma$, may 
be determined by studying $e^+e^-\to Hf\bar{f}$ processes at future $e^+e^-$
linear colliders.
By using the optimal-observable method, which makes use of all available
experimental information, we find out which combinations of the various $HZV$
coupling terms may be constrained most efficiently with high luminosity.
We also assess the benefits of measuring the tau-lepton helicities, 
identifying the bottom-hadron charges, polarizing the electron beam and
running at two different collider energies.
The $HZZ$ couplings are generally found to be well constrained, even without
these options, while the $HZ\gamma$ couplings are not.
The constraints on the latter may be significantly improved by beam
polarization.
\end{abstract}

\newpage

\section{Introduction}
\cleqn

The standard model (SM) of elementary-particle physics predicts a neutral 
scalar Higgs boson $H$ as a remnant of the spontaneous breaking of its gauge
symmetry.
This particle is the only undiscovered ingredient of the SM so far.
The experiments at the CERN Large Electron-Positron Collider (LEP2) were able
to place lower bounds on its mass in the range 91.0--98.8~GeV at the 95\%
confidence level (CL) \cite{Nielsen:1999ct}.
The search for the Higgs boson is a prime target of future colliders.
Once the Higgs boson is found, its properties and interactions with other
particles may be studied in detail with $e^+e^-$ linear colliders.
If the Higgs boson is light, the bremsstrahlung process $e^+e^- \to \h Z$
is expected to be the most promising process to study its properties and
interactions and to search for deviations from the SM predictions.

The purpose of this paper is to study systematically 
the sensitivities to general, non-standard couplings
among the Higgs boson, the $Z$ boson and a neutral vector boson $V$
($V=Z,\gamma$).
Since the $Z$ boson has spin one, we take into account the angular
distributions of its subsequent decays to fermion-antifermion pairs, in order
not to loose information on the interference between amplitudes with different
$Z$-boson helicities.
On the other hand, we treat the Higgs boson as a final-state particle because
it has spin zero.
Thus, we study the production and decay processes
\[
       e^+e^- \to \h Z ; \qquad Z \to f\bar{f}
\]
to obtain sensitivity to general $HZV$ couplings.

We first review previous studies on related problems.
The angular distribution of $e^+e^- \to \h f\bar{f}$
has been analyzed for the SM at the tree level in
\cite{Kelly:1981ue}.
Expressions for the cross sections have been elaborated for beam polarization
in \cite{Romao:1987cz}.
Radiative corrections have been investigated in \cite{Fleischer:1983af}.
A comprehensive review of the Higgs boson properties has been given in
\cite{Kniehl:1994ay}.
The $HZZ$ form factors have been introduced in the study of composite light
Higgs bosons \cite{Godbole:1983mx}.
Effects of the non-standard couplings have been discussed
in \cite{Rattazzi:1988ye,Hagiwara:1994sw}.
$Z$-boson decay angular distribution in the process $e^+e^-\to \h Z$ have been
analyzed as a means of distinguishing a scalar form a pseudoscalar Higgs boson
in \cite{Barger:1994wt}.

We employ the optimal-observable method
\cite{Atwood:1992ka,Davier:1993nw,Diehl:1994br,Gunion:1996vv} to obtain
constraints on the $HZV$ couplings.
This method provides the most efficient way to extract physical parameters
from experimental data in the sense that the statistical errors on these
parameters are minimized.
Atwood and Soni have introduced optimal observable quantities in their
analysis of electromagnetic form factors of the top quark \cite{Atwood:1992ka}.
The optimal-observable method has also been used in the measurement of the tau
polarization \cite{Davier:1993nw}.
This method has then been extended to the many-parameter case.
It has been applied to the determination of the electroweak triple-gauge-boson
\cite{Diehl:1994br}, $Ht\bar{t}$ and $HZZ$ couplings \cite{Gunion:1996vv}.

This paper is organized as follows.
In the next section, we cast the differential cross sections of the process
\mbox{$e^+e^- \to \h f\bar{f}$} into a compact form, to which the
optimal-observable method can be applied.
We then discuss the properties of the various terms therein under discrete
symmetries.
In Sect.~\ref{sec:Helicity}, we introduce an effective Lagrangian for the
$HZV$ interactions and calculate the helicity amplitudes of the process
\mbox{$e^+e^- \to \h Z$}.
In Sect.~\ref{sec:optimal}, we determine the achievable errors on the optimal
observables introduced in the general expansion of the cross section.
In Sect.~\ref{sec:numerical}, the optimal constraints on the effective
coupling constants are discussed for typical experimental situations.
Our conclusions are presented in Sect.~\ref{sec:conclusion}.

\boldmath
\section{Cross section of $e^+e^- \to \h f\bar{f}$}
\unboldmath
\cleqn
\label{sec:dcs}

In this section, we present the general angular distributions of the
differential cross section of the production and decay process,
\bmeq
\label{eq:eezff}
 e^-\left(p_e^{},\frac{\hi}{2}\right)
 + e^+\left(p_{\bar{e}}^{},-\frac{\hi}{2}\right)
  &\to& Z^*/\gamma^*(\pv)
  \to \h(p_\h^{}) + Z(p_Z^{},\lambda)  , \label{eq:eezh}\\
 Z(p_Z^{},\lambda) &\to&
  f\left(p_f^{},\frac{\hf}{2}\right) 
 + \bar{f}\left(p_{\bar{f}}^{},-\frac{\hf}{2}\right),
 \label{eq:zff}
\emeq
in a compact form suitable for the optimal-observable method.
The four-momentum and helicity of each particle is indicated in parentheses;
we have $\hi=\pm1$, $\hf=\pm1$ and $\lambda=0, \pm1$ .

We evaluate the production process (\ref{eq:eezh}) in the centre-of-mass (CM)
frame of the colliding beams.
The production amplitude is then a function of the scattering angle $\Theta$
enclosed between the incoming electron and outgoing $Z$-boson
three-momenta, ${\bf p}_e$ and
${\bf p}_Z$, respectively.
The $Z$-boson helicity $\lambda$ is defined in the CM frame of the
colliding beams.
The $y$ axis is chosen along the ${\bf p}_e \times {\bf p}_Z$ direction.
The decay process (\ref{eq:zff}) is described in the rest frame of the
outgoing $Z$ boson.
Here, the $z$-axis is chosen along the direction of ${\bf p}_Z$ (before the 
boost).
The decay amplitude is a function of the polar angle $\theta$ and the
azimuthal angle $\varphi$ of the $f$ three-momentum.

\subsection{Angular distributions}

The angular distributions of the differential cross section of
$e^+e^- \to \h f\bar{f}$ may be written as
\begin{equation}
 \frac{d\sigma}{d\cos\Theta d\cos\theta d\varphi}
 = \sum_{i=1}^{9}
 \left[
  c^{(V)}_i F^{(V)}_i(\Theta, \theta, \varphi )
  + c^{(A)}_i F^{(A)}_i(\Theta, \theta, \varphi )
	      \right],
\end{equation}
where $c_i^{(V,A)}$ are model-dependent coefficients and $F_i^{(V,A)}$ are
known functions of the angles $\Theta$, $\theta$ and $\varphi$.
We shall present the definitions of $c_i^{(V,A)}$ and $F_i^{(V,A)}$ below.
The functions $F_i^{(V,A)}$ depend on the flavor of the final-state fermion
$f$ and the polarization $P$ of the initial-state electron;
we have $P=\pm1$ if the electron beam is purely right/left-handed.
We assume that the positron beam is unpolarized.
In the derivation of $c_i^{(V,A)}$ and $F_i^{(V,A)}$, we use the narrow-width
approximation for the $Z$-boson propagator and the SM amplitude for the decay
process $Z \to f\bar{f}$.

We define reduced helicity amplitudes $\hat{M}^\lambda_\hi$ by extracting the
angular dependence from the helicity amplitudes ${M}^\lambda_\hi$ for
$e^+e^-\to \h Z$ as
\begin{equation}
\label{eq:reduce}
 {M}^\lambda_\hi (e^+ e^- \to \h Z) =
  \hat{M}^\lambda_\hi
  d^1_{\hi,\lambda}(\Theta) ,
\end{equation}
where
\begin{eqnarray}
  d^1_{\hi,\lambda=0}(\Theta) = - \frac1{\sqrt{2}}\hi \sin\Theta , \qquad
  d^1_{\hi,\lambda=\pm}(\Theta) = \frac12(1 + \hi\lambda \cos\Theta) .
\end{eqnarray}
The amplitudes $\hat{M}^\lambda_\hi$ do not depend on $\Theta$.

The coefficients $c_i^{(V)}$ and $c_i^{(A)}$ are expressed in terms of the
amplitudes $\hat{M}^\lambda_\hi$ as
\bmeq
\label{eq:ciVA}
 c^{(V,A)}_1 &=&   \MMaR0 \pm \MMaL0 , \\
 c^{(V,A)}_2 &=&  \MMaR+ + \MMaR-
  \pm \left( \MMaL+ + \MMaL- \right) , \\
 c^{(V,A)}_3 &=&    \Re \left[\MMR0+ + \MMR-0\right]
  \pm \Re\left[ \MML0+ + \MML-0\right] , \\
 c^{(V,A)}_4 &=&   \Re\left[\MMR-+\right] \pm \Re\left[\MML-+\right] , \\
 c^{(V,A)}_5 &=&  \Im\left[\MMR0+ + \MMR-0\right]
  \pm \Im\left[\MML0+ + \MML-0\right], \\
 c^{(V,A)}_6 &=&  \Im\left[\MMR-+\right] \pm \Im\left[\MML-+\right] , \\
 c^{(V,A)}_7 &=& \MMaR+ - \MMaR- \pm \left(\MMaL+ - \MMaL-\right), \\
 c^{(V,A)}_8 &=&  \Re\left[\MMR0+ - \MMR-0\right]
  \pm \Re\left[\MML0+ - \MML-0\right], \\
 c^{(V,A)}_9 &=&  \Im\left[\MMR0+ - \MMR-0\right]
  \pm \Im\left[\MML0+ - \MML-0\right].
\emeq
Here, the $+$ ($-$) sign refers to $V$ ($A$),
and the subscript $R$ ($L$) stands for $\hi=+1$ ($\hi=-1$).
These eighteen coefficients contain all observable consequences of the reduced amplitudes.

The functions $F_i^{(V)}$ are defined as
\bmeq
\label{eq:FiV}
 F^{(V)}_1 &=& \frac{r}{4} \sin^2\Theta \sin^2\theta , \\
 F^{(V)}_2 &=& \frac{r}{16}(1+\cos^2\Theta)(1+\cos^2\theta) 
    -   \frac{r P A_f}{4} \cos\Theta \cos\theta , \\
 F^{(V)}_3 &=& - \frac{r}{16} \sin2\Theta \sin2\theta \cos\varphi 
    +   \frac{r P A_f}{4}\sin\Theta \sin\theta \cos\varphi , \\
 F^{(V)}_4 &=& \frac{r}{8} \sin^2\Theta \sin^2\theta \cos2\varphi , \\
 F^{(V)}_5 &=&  - \frac{r}{16}\sin2\Theta \sin2\theta \sin\varphi
    +   \frac{r P A_f}{4}\sin\Theta \sin\theta \sin\varphi , \\
 F^{(V)}_6 &=&  \frac{r}{8} \sin^2\Theta \sin^2\theta \sin2\varphi , \\
 F^{(V)}_{7} &=&  - \frac{r A_f}{8} (1+\cos^2\Theta)\cos\theta 
    +  \frac{r P}{8} \cos\Theta (1+\cos^2\theta) , \\
 F^{(V)}_8 &=&  \frac{r A_f}{8}\sin2\Theta \sin\theta \cos\varphi 
    -  \frac{r P}{8}\sin\Theta \sin2\theta \cos\varphi , \\
 F^{(V)}_9 &=& \frac{r A_f}{8}\sin2\Theta \sin\theta \sin\varphi 
    - \frac{r P}{8}\sin\Theta \sin2\theta \sin\varphi .
\emeq
The common coefficient $r$ contains some phase space factors and the branching
fraction of the $Z\to f\bar f$ decay,
\begin{equation}
 \label{eq:Br}
r=\frac14\,\frac{\beta_{\h Z}}{32\pi s}\,\frac{3}{4\pi}
\mbox{Br}\left(Z\to f\bar{f}\right),
\end{equation}
and $A_f$ is the left-right asymmetry of this decay,
\begin{equation}
 \label{eq:Af}
A_f=\frac{\left(\gfL\right)^2-\left(\gfR\right)^2}
{\left(\gfL\right)^2+\left(\gfR\right)^2}.
\end{equation}
Here, $s$ is the square of the CM energy, and $\beta_{\h Z}$ is the two-body
phase-space factor,
\begin{equation}
 \beta_{\h Z} = \sqrt{1 - 2\frac{\mmz+\mmh}{s}
  + \left(\frac{\mmz - \mmh }{s}\right)^2} .
\end{equation}
The functions $F^{(A)}_i$ are obtained by flipping the $P$ dependence.
Specifically, if we write $F_i^{(V)}=\alpha_i + \beta_i P$, with
$P$-independent functions $\alpha_i$ and $\beta_i$, then we have
$F_i^{(A)}=\alpha_i P + \beta_i$.
The explicit expressions read
\bmeq
\label{eq:FiA}
 F^{(A)}_1 &=& \frac{r P}{4} \sin^2\Theta \sin^2\theta , \\
 F^{(A)}_2 &=& \frac{r P}{16}(1+\cos^2\Theta)(1+\cos^2\theta) 
    -   \frac{r A_f}{4} \cos\Theta \cos\theta , \\
 F^{(A)}_3 &=& - \frac{r P}{16}\sin2\Theta \sin2\theta \cos\varphi 
    +   \frac{r A_f}{4}\sin\Theta \sin\theta \cos\varphi , \\
 F^{(A)}_4 &=& \frac{r P}{8} \sin^2\Theta \sin^2\theta \cos2\varphi , \\
 F^{(A)}_5 &=&  - \frac{r P}{16}\sin2\Theta \sin2\theta \sin\varphi
    +   \frac{r A_f}{4}\sin\Theta \sin\theta \sin\varphi , \\
 F^{(A)}_6 &=&  \frac{r P}{8} \sin^2\Theta \sin^2\theta \sin2\varphi , \\
 F^{(A)}_{7} &=&  - \frac{r P A_f}{8} (1+\cos^2\Theta)\cos\theta 
    +  \frac{r}{8} \cos\Theta (1+\cos^2\theta) , \\
 F^{(A)}_8 &=&   \frac{r P A_f}{8}\sin2\Theta \sin\theta \cos\varphi 
    -  \frac{r}{8}\sin\Theta \sin2\theta \cos\varphi , \\
 F^{(A)}_9 &=&  \frac{r P A_f}{8}\sin2\Theta \sin\theta \sin\varphi 
    - \frac{r}{8}\sin\Theta \sin2\theta \sin\varphi .
\emeq
The three angular functions $F^{(A)}_1$, $F^{(A)}_4$ and $F^{(A)}_6$ vanish
if $P=0$.
One could measure $c^{(A)}_1$, $c^{(A)}_4$ and $c^{(A)}_6$ 
if $|P|\neq 0$ by combining
experiments with opposite polarizations $P=|P|$ and $P=-|P|$.

For most of the hadronic decay modes of the $Z$ boson, the final-state
fermions $f$ and $\bar{f}$ cannot be distinguished.
Then, we have to average over the configurations with
$(\Theta, \theta, \varphi)$ and $(\Theta, \pi-\theta, \varphi \pm\pi)$ as
\begin{equation}
\bar F^{(V,A)}_i (\Theta, \theta, \varphi)
= \frac12\left[ F^{(V,A)}_i(\Theta, \theta, \varphi)
 + F^{(V,A)}_i(\Theta, \pi-\theta, \varphi \pm\pi)\right] .
\end{equation}
This corresponds to setting $A_f=0$ if $f$ is a quark.
If $P=0$, then one can measure the coefficients
$c_1^{(V)}, \dots, c_6^{(V)}$ and $c_7^{(A)}, \dots, c_9^{(A)}$,
while the coefficients $c_7^{(V)}, \dots, c_9^{(V)}$ and
$c_1^{(A)}, \dots, c_6^{(A)}$ are only measurable if $P\neq 0$.

When the $Z$ bosons decay to neutrino pairs, one can only measure the $\Theta$
distribution, so that the $(\theta,\varphi)$ dependences should be integrated
out.
Then, we have
\begin{equation}
 \frac{d\sigma}{d\cos\Theta}
  = \sum_{i=1,2,7}\left[c^{(V)}_i \tilde F^{(V)}_i(\Theta)
  + c^{(A)}_i \tilde F^{(A)}_i(\Theta)\right],
\end{equation}
where
\bmeq
 \tilde F^{(V)}_1 &=& \frac{2\pi r}{3} \sin^2\Theta  , \\
 \tilde F^{(V)}_2 &=& \frac{\pi r}{3} (1+\cos^2\Theta) , \\
 \tilde F^{(V)}_7 &=& \frac{2\pi r P}{3} \cos\Theta , \\
 \tilde F^{(A)}_1 &=& \frac{2\pi r P}{3} \sin^2\Theta  , \\
 \tilde F^{(A)}_2 &=& \frac{\pi r P}{3} (1+\cos^2\Theta) , \\
 \tilde F^{(A)}_7 &=& \frac{2\pi r}{3} \cos\Theta ,
\emeq
while the other $\tilde F^{(V,A)}_i$ functions vanish.

\subsection{Discrete symmetries}
\label{sec:cpt}

We now discuss the properties of the $F_i^{(V,A)}$ functions under the
discrete symmetries $CP$ and $CP\tilde T$.
Here, $\tilde T$ is the {\it naive} time reversal symmetry, which flips
the momentum and spin of all the particles,
but does not reverse the time flow from the initial state
to the final state.
The non-vanishing of the $CP\tilde T$-odd coefficients is related to the
presence of absorptive parts in the amplitudes \cite{Hagiwara:1987vm}.

The electron beam polarization $P$, the decay asymmetry $A_f$ and the angular
variables transform under the discrete symmetries as
\bmeq
  ( P , A_f  ; \Theta, \theta, \varphi)
  &\stackrel{CP}{\to}& ( P , A_f ; \pi-\Theta, \pi-\theta, 2\pi - \varphi) ,
   \\
  ( P , A_f  ; \Theta, \theta, \varphi)
  &\stackrel{\tilde{T}}{\to}& ( P , A_f ; \Theta, \theta, 2\pi-\varphi) .
\emeq
We can then obtain the symmetry properties of the $F^{(V,A)}_i$ functions,
which are summarized in Table~\ref{table:FiVA}.

\begin{table}[phtb]
 \begin{center}
  \caption{$CP$ and $CP\tilde T$ properties of the $F^{(V,A)}_i$ functions.
A $+$ ($-$) sign means even (odd) under the symmetry.} \label{table:FiVA}
  \begin{tabular}[tb]
   {c|ccccccccc}
   \hdick \\[-2.0ex]
   $i$ & 1& 2& 3& 4& 5& 6& 7& 8& 9 \\
   \hline
   $CP$ &          $+$& $+$& $+$& $+$& $-$& $-$& $-$& $-$& $+$ \\
   $CP\tilde{T}$& $+$& $+$& $+$& $+$& $+$& $+$& $-$& $-$& $-$ \\
   \hline
  \end{tabular}
 \end{center}
\end{table}

The coefficients $c^{(V,A)}_i$ have the same symmetry properties as the
functions $F^{(V,A)}_i$.
The coefficients $c^{(V,A)}_5$ and $c^{(V,A)}_6$ are sensitive to $CP$-odd and
$CP\tilde{T}$-even quantities.
When the new-physics effects are generated by exchanges of heavy particles, 
then the induced vertices should be $CP\tilde{T}$ even.
The three $CP\tilde{T}$-odd coefficients $c_7^{(V,A)}$, $c_8^{(V,A)}$ and
$c_9^{(V,A)}$ should be proportional to the absorptive parts of the amplitudes
which contain light particles in the loops.

\boldmath
\section{Helicity amplitudes for $e^+e^- \to\h Z$}
\unboldmath
\cleqn
\label{sec:Helicity}

In this section, we first introduce general couplings and effective form
factors for the $\h ZZ$ and $\h Z\gamma$ interactions.
We then present the helicity amplitudes of $e^+e^-\to\h Z$ using these form
factors.

We adopt the effective $\h ZV$ interaction Lagrangian from
\cite{Hagiwara:1994sw}.
It reads
\begin{eqnarray}
 {\cal L}_{\rm eff} &=& (1+a_Z) \frac{\gZ\mz}{2} \h Z_\mu Z^\mu
  + \frac{\gZ}{\mz} \sum_{V=Z,\gamma}
  \left[
   b_V \h Z_{\mu\nu}V^{\mu\nu} 
   \vphantom{\partial_\mu\h \tilde{b}_V}
    \right.
    \nonumber \\
 &&{}+\left.
   c_V \left(\partial_\mu\h Z_\nu - \partial_\nu\h Z_\mu\right) V^{\mu\nu}
   + \tilde{b}_V \h Z_{\mu\nu}\tilde{V}^{\mu\nu}
	\right],
\label{eq:lag}
\end{eqnarray}
where $V_{\mu\nu}= \partial_\mu V_\nu - \partial_\nu V_\mu$ and
$\tilde{V}_{\mu\nu} = \epsilon_{\mu\nu\alpha\beta}V^{\alpha\beta}$
with the convention $\epsilon_{0123}=+1$.
We have neglected the scalar component of the vector bosons by putting
\begin{equation}
\partial_\mu Z^\mu = \partial_\mu V^\mu = 0 .
\label{eq:scalar}
\end{equation}
Then, the most general parameterization of the $\h ZV$ interaction
involves seven couplings, $a_Z$, $b_Z$, $c_Z$, $b_\gamma$, $c_\gamma$,
$\tilde{b}_Z$ and $\tilde{b}_\gamma$,
which are constants as long as we only consider operators through mass
dimension five.
We note in particular that the operator identity
\begin{eqnarray}
 \h Z_\mu \partial^2 Z^\mu
  &=& \h Z_\mu \partial_\nu Z^{\nu\mu}
  \nonumber \\
 &=& - \frac12 \h Z_{\mu\nu}Z^{\mu\nu}
  - \frac12 (\partial_\mu\h Z_\nu - \partial_\nu\h Z_\mu) Z^{\mu\nu}
\label{eq:Op.Identity}
\end{eqnarray}
holds under the condition (\ref{eq:scalar}).
The five couplings $a_Z$, $b_Z$, $c_Z$, $b_\gamma$ and $c_\gamma$ are $CP$
even, while the remaining two couplings, $\tilde{b}_Z$ and $\tilde{b}_\gamma$,
are $CP$ odd.
In the effective Lagrangian (\ref{eq:lag}), we have factored out the $Z$-boson
coupling $g_Z^{}$ and appropriate powers of $\mz$ to render the couplings
dimensionless.
In the SM, we have $a_Z= b_V=c_V=\tilde{b}_V=0$ at the tree level.

The form factors for the generic $\h Z_\alpha V_\beta$ vertex may then be
written as
\begin{equation}
 \Gamma^V_{\alpha\beta}(\pv,p_Z^{}) = \gZ \mz
  \left[
   h^V_1(s) g_{\alpha\beta}
   +  \frac{h^V_2(s)}{\mmz} \pv_\alpha p_{Z\beta}^{}
   +  \frac{h^V_3(s)}{\mmz} \epsilon_{\alpha\beta\mu\nu} \pv^\mu p_Z^\nu
   \right] ,
\end{equation}
where the virtual $V$-boson momentum $\pv$ is taken to be incoming and the
$Z$-boson momentum $p_Z^{}$ to be outgoing, as depicted in
Figure~\ref{fign:pflow} and process (\ref{eq:eezff}), and $s=\pv^2$.
All form factors $h_i^V$ are dimensionless functions of $s$.
The four form factors $h^Z_1$, $h^Z_2$, $h^\gamma_1$ and $h^\gamma_2$ are $CP$
even, while the two form factors $h^Z_3$ and $h^\gamma_3$ are $CP$ odd. 
It is straightforward to express the form factors in terms of the seven
couplings of the effective Lagrangian (\ref{eq:lag}):
\bmeq \label{eq:formfactor}
h^Z_1(s) &=& (1+a_Z) + 2c_Z\frac{s + \mmz}{\mmz}
 + 2(b_Z-c_Z)\frac{s + \mmz - \mmh}{\mmz}, \\
h^Z_2(s) &=& - 4 (b_Z - c_Z) , \\
h^Z_3(s) &=& - 4 \tilde{b}_Z , \\
h^\gamma_1(s) &=& 2 c_\gamma \frac{s}{\mmz}
 + (b_\gamma - c_\gamma) \frac{s + \mmz - \mmh}{\mmz}, \\
h^\gamma_2(s) &=& - 2 (b_\gamma - c_\gamma) , \\
h^\gamma_3(s) &=& - 2 \tilde{b}_\gamma .
\emeq
Although the effective Lagrangian has seven couplings, there are only six form
factors.
Thus, one combination of couplings cannot be measured at one given collider
energy.
Details will be discussed in Sect.~\ref{sec:numerical}.

We now evaluate the helicity amplitudes $M^\lambda_\hi(e^+e^-\to\h Z)$
for the production process.
After extracting the angular dependence according to (\ref{eq:reduce}), we
obtain the reduced amplitudes\footnote{%
In \cite{Hagiwara:1994sw}, there is a misprint in the relative sign of the
helicity amplitudes.
There should be an overall minus sign on the right-hand side of (2.3) therein.
Furthermore,
$\Im [(\hat{M}_\sigma^+ + \hat{M}_\sigma^-)(\hat{M}_\sigma^0)^*]$
and
$\Im [(\hat{M}_\sigma^+ - \hat{M}_\sigma^-)(\hat{M}_\sigma^0)^*]$
should be interchanged in the first column of Table~1.}
as functions of $s$:
\bmeq
\label{eq:amp}
 \hat{M}^{\lambda=0}_\hi(s)
  &=& 
  - g_Z^2 \ges \sqrt{2s} E_Z D_Z(s)
  \left(h_1^Z + h_2^Z \frac{\sqrt{s} E_Z \beta_Z^2}{m_Z^2}  \right)
    \nonumber \\
   &&  \mbox{}
  + e g_Z^{} \sqrt{2s} E_Z D_\gamma(s)
  \left(h_1^\gamma 
  + h_2^\gamma \frac{\sqrt{s} E_Z \beta_Z^2}{m_Z^2} \right)
  , \\
 \hat{M}^{\lambda=\pm}_\hi(s)
  &=& 
  - g_Z^2 \ges \sqrt{2s} \mz D_Z(s)
  \left(h_1^Z + i\lambda h_3^Z \frac{\sqrt{s} E_Z \beta_Z}{\mmz} \right)
    \nonumber \\
   &&  \mbox{}
  + e g_Z^{} \sqrt{2s} \mz D_\gamma(s)
  \left(h_1^\gamma + i\lambda h_3^\gamma \frac{\sqrt{s} E_Z \beta_Z}{\mmz}
 \right) .
\emeq
where $\geR = \sin^2\theta_W$, $\geL = -1/2 + \sin^2\theta_W$,
$\beta_Z = \sqrt{1 - \mmz/E_Z^2}$ and
\begin{eqnarray}
 D_Z(s) &=& \frac1{s - \mmz + i \mz\Gamma_Z} , \\
 D_\gamma(s) &=& \frac1{s} .
\end{eqnarray}
The energy of the outgoing $Z$ boson is
\begin{equation}
E_Z = \frac{\sqrt{s}}{2}\left(1 - \frac{\mmh  - \mmz}{s}\right).
\end{equation}
The couplings $\geR$ and $\geL$ have almost the same magnitudes, but their
signs are opposite to each other.
Thus, the coefficients $c_i^{(V)}$ are sensitive to the $\h ZZ$ couplings,
while the coefficients $c_i^{(A)}$ are sensitive to the $\h Z\gamma$ couplings.

\section{Optimal observables}
\cleqn
\label{sec:optimal}

We now employ the optimal-observable method
\cite{Diehl:1994br,Gunion:1996vv}
to obtain the errors on the coefficients $c_i^{(V)}$ and $c_i^{(A)}$.
In order to simplify the notation, let us re-sequence the coefficients and
functions for the time being as
\bmeq
(c_1, \ldots, c_{18})
&=& (c_1^{(V)}, \ldots, c_9^{(V)}, c_1^{(A)}, \ldots, c_9^{(A)}) , \\
(F_1, \ldots, F_{18})
&=& (F_1^{(V)}, \ldots, F_9^{(V)}, F_1^{(A)}, \ldots, F_9^{(A)}) .
\emeq
According to the optimal-observable method, the covariance matrix $V_{ij}$ for
the coefficients $c_i$ is given by
\begin{equation}
\label{eq:Vij}
 V_{ij}^{-1} = L \int \frac
  {F_i(\Theta, \theta, \varphi) F_i(\Theta, \theta, \varphi)}
  {\Sigma_{\mathrm{SM}}(\Theta, \theta, \varphi)}
  d\cos\Theta \cos\theta d\varphi ,
\end{equation}
where $L$ is the integrated luminosity of the experiment and
\begin{equation}
 \Sigma_{\mathrm{SM}}(\Theta, \theta, \varphi)
  = \frac{d\sigma_{\mathrm{SM}}}
  {d\cos\Theta \cos\theta d\varphi}(\Theta, \theta, \varphi) .
\end{equation}
The statistical error on $c_i$ is $\sqrt{V_{ii}}$,
and the correlation between the error on $c_i$ and that on $c_j$
is $V_{ij}/\sqrt{V_{ii} V_{jj}}$.
We use this method to obtain $V^{-1}_{ij}$ for each decay mode of the
$Z$ boson.  We then combine these results for all the $Z$-boson decay
modes.

The differential cross section $\Sigma_{\mathrm{SM}}$ is a linear combination
of $F_1^{(V)},\ldots, F_4^{(V)}$ and $F_1^{(A)},\ldots, F_4^{(A)}$,
which are $CP$ and $CP\tilde{T}$ even.
The component $V_{ij}$ of the covariance matrix vanishes if $F_i$ and $F_j$
have different $CP$ or $CP\tilde{T}$ properties.
Thus, the covariance matrix is block diagonalized into four sub-matrices
according to the $CP$ and $CP\tilde{T}$ properties of the $F_i$ functions 
discussed in Sect.~\ref{sec:cpt}.
Notice that this argument is only valid if we integrate in (\ref{eq:Vij})
over the full angle domains.
In practice, there are excluded regions due to the geometry of the detectors
or cuts for selecting events.
Thus, the block diagonal structure of the covariance matrix is only 
approximately realized in practice.
In the present study, we shall integrate over the full phase space.

We estimate how the optimal errors are reduced by the following three
additional techniques.
The first one is the tau helicity measurement.
We adopt $\efft = 40\%$ as the efficiency factor to determine the
helicities of the decaying $\tau^+$ or $\tau^-$ leptons.
The second technique is the electric-charge identification for the bottom
quarks and antiquarks.
The charge of a hadron $B$ containing one $b$ or $\bar{b}$ quark can be
identified via the decay mode $B\to l\nu+X$.
We assume an efficiency of $\effb = 20\%$ for identifying the charges of the
decaying $b$ or $\bar{b}$ hadrons.
The third technique is to employ electron beam polarization $P$.
We take $|P| = 90\%$ as the target polarization.
Specifically, we assume that one half of the beam is polarized with $P=0.9$
and the other half with $P=-0.9$.

For the fraction $\efft$ of the $Z\to\tau^+\tau^-$ decays, one can
distinguish the tau polarization.
In order to assess the possible benefits of the tau polarization measurement,
we make the simple assumption that the efficiency for observing a left- or
right-handed tau lepton is $\efft$.
Then, we substitute in (\ref{eq:FiV}) and (\ref{eq:FiA})
\bmeq
 r &=& \frac{3\beta_{\h Z}}{512\pi^2 s}
  \mbox{Br}(Z\to \tau^+ \tau^-) 
    \frac{\left(g^{\tau}_{L/R}\right)^2}
	{\left(g^{\tau}_L\right)^2+\left(g^{\tau}_R\right)^2}
    \efft, \\
 A_f &=& \pm1 
\emeq
for left/right-handed tau leptons.
In actual experiments, one should not only estimate the efficiency factor
$\efft$ for each pair of $\tau^+$ and $\tau^-$ decay modes, but also a
correlation between the constraints from $\tau^-_R$ production and those from
$\tau^-_L$ production.
We return to this problem at the end of this section.

Throughout our numerical analysis, we set
$\mz=91.187$~GeV \cite{Caso:1998tx},
$\alpha=1/128.9$ \cite{Eidelman:1995ny},
$\sin^2{\theta_W^{}}=0.2312$
and
$g_Z^{} = \sqrt{4\sqrt{2} G_F^{} \mmz } = 0.74070$.   
As an example, we show results for the Higgs boson mass $m_\h = 120$~GeV\@,
the CM energy $\sqrt{s} = 250$~GeV and the nominal integrated
luminosity $L = 10$~fb$^{-1}$.

We first present the results for $\efft=\effb=P=0$.
The results for the $CP$-even and $CP\tilde{T}$-even coefficients and their
block in the covariance matrix are
\begin{equation}
\begin{array}[tb]{ccc}
 c_1^{(V)} &=& .0208 \pm  .0011 \\
 c_2^{(V)} &=& .0271 \pm  .0012 \\
 c_3^{(V)} &=& .0336 \pm  .0049 \\
 c_4^{(V)} &=& .0136 \pm  .0026 \\
 c_2^{(A)} &=& -.004 \pm  .025 \\
 c_3^{(A)} &=& -.005 \pm  .020
\end{array},
 \qquad
 \left(
  \begin{array}[tb]{rrrrrr}
  \one & & & & & \\
  -.66 & \one & & & & \\
  -.02 &  .13 & \one & & & \\
   .10 & -.02 &  .06 & \one & & \\
   .00 & -.00 & -.00 & -.00 & \one & \\
  -.00 & -.00 & -.00 & -.00 & .10 & \one
  \end{array}
  \right) .
\end{equation}
The results for the $CP$-odd and $CP\tilde{T}$-even coefficients are
\begin{equation}
\begin{array}[tb]{ccl}
 c_5^{(V)} &=& 0 \pm .0047 \\
 c_6^{(V)} &=& 0 \pm .0026 \\
 c_5^{(A)} &=& 0 \pm .018
\end{array},
 \qquad
 \left(
  \begin{array}[tb]{rrr}
  \one & & \\
   .07 & \one & \\
  -.00 & -.00 & \one
  \end{array}
  \right) .
\end{equation}
The results for the $CP$-odd and $CP\tilde{T}$-odd coefficients are
\begin{equation}
\begin{array}[tb]{ccl}
 c_7^{(V)} &=& 0 \pm  .021 \\
 c_8^{(V)} &=& 0 \pm  .042 \\
 c_7^{(A)} &=& 0 \pm  .0010 \\
 c_8^{(A)} &=& 0 \pm  .0022 
\end{array},
 \qquad
 \left(
  \begin{array}[tb]{rrrr}
  \one & & & \\
   .12 & \one & & \\
  -.00 & -.00 & \one & \\
  -.00 & -.00 &  .11 & \one
  \end{array}
  \right) .
\end{equation}
The results for the $CP$-even and $CP\tilde{T}$-odd coefficients are
\begin{equation}
\begin{array}[tb]{ccl}
 c_9^{(V)} &=& 0 \pm  .039 \\
 c_9^{(A)} &=& 0 \pm  .0021 
\end{array},
 \qquad
 \left(
  \begin{array}[tb]{rr}
  \one & \\
  -.00 & \one
  \end{array}
  \right) .
\end{equation}
There are no constraints on $c_1^{(A)}$, $c_4^{(A)}$ and $c_6^{(A)}$ because
$F_1^{(A)}$, $F_4^{(A)}$ and $F_6^{(A)}$ vanish if $P=0$.
The errors on $c^{(A)}_2$, $c^{(A)}_3$, $c^{(A)}_5$, $c^{(V)}_7$, $c^{(V)}_8$
and $c^{(V)}_9$ are relatively large because the corresponding $F_i^{(V,A)}$
functions in (\ref{eq:FiA}) are suppressed by the smallness of $A_f$ and the
vanishing of $P$.

Next, we present the results for $\efft = 40\%$, $\effb = 20\%$ and
$|P| = 90\%$.
The results for the $CP$-even and $CP\tilde{T}$-even coefficients are
\begin{equation}
\begin{array}[tb]{ccc}
 c_1^{(V)} &=& .0208 \pm  .0011 \\
 c_2^{(V)} &=& .0271 \pm  .0012 \\
 c_3^{(V)} &=& .0336 \pm  .0028 \\
 c_4^{(V)} &=& .0136 \pm  .0026 \\
 c_1^{(A)} &=& -.0031 \pm  .0012 \\
 c_2^{(A)} &=& -.0041 \pm  .0013 \\
 c_3^{(A)} &=& -.0050 \pm  .0027 \\
 c_4^{(A)} &=& -.0020 \pm  .0029
\end{array},
 \qquad
 \left(
  \begin{array}[tb]{rrrrrrrr}
  \one & & & & & & & \\
  -.65 & \one & & & & & & \\
   .14 &  .06 & \one & & & & & \\
   .10 & -.01 &  .20 & \one & & & & \\
  -.13 &  .08 & -.03 & -.01 & \one & & & \\
   .08 & -.13 &  .00 & -.00 & -.65 & \one & & \\
  -.03 &  .00 & -.06 & -.04 &  .15 &  .05 & \one & \\
  -.01 & -.00 & -.04 & -.13 &  .10 & -.01 &  .20 & \one
  \end{array}
  \right) .
\end{equation}
The results for the $CP$-odd and $CP\tilde{T}$-even coefficients are
\begin{equation}
\begin{array}[tb]{ccl}
 c_5^{(V)} &=& 0 \pm  .0031 \\
 c_6^{(V)} &=& 0 \pm  .0026 \\
 c_5^{(A)} &=& 0 \pm  .0030 \\
 c_6^{(A)} &=& 0 \pm  .0029
\end{array},
 \qquad
 \left(
  \begin{array}[tb]{rrrr}
  \one & & & \\
   .18 & \one & & \\
  -.11 & -.04 & \one & \\
  -.03 & -.13 &  .19 & \one
  \end{array}
  \right) .
\end{equation}
The results for the $CP$-odd and $CP\tilde{T}$-odd coefficients are
\begin{equation}
\begin{array}[tb]{ccl}
 c_7^{(V)} &=& 0 \pm  .0011 \\
 c_8^{(V)} &=& 0 \pm  .0023 \\
 c_7^{(A)} &=& 0 \pm  .0010 \\
 c_8^{(A)} &=& 0 \pm  .0021
\end{array},
 \qquad
 \left(
  \begin{array}[tb]{rrrr}
  \one & & & \\
   .17 & \one & & \\
  -.13 & -.03 & \one & \\
  -.03 & -.12 &  .16 & \one
  \end{array}
  \right) .
\end{equation}
The results for the $CP$-even and $CP\tilde{T}$-odd coefficients are
\begin{equation}
\begin{array}[tb]{ccl}
 c_9^{(V)} &=& 0 \pm  .0023 \\
 c_9^{(A)} &=& 0 \pm  .0021
\end{array},
 \qquad
 \left(
  \begin{array}[tb]{rr}
  \one & \\
  -.13 & \one
  \end{array}
  \right) .
\end{equation}
The errors on $c_i^{(A)}$ with $i=1,\ldots,6$ and $c_i^{(V)}$ with $i=7,8,9$
are reduced to the level of those on the other coefficients because the
suppression of the corresponding $F_i^{(V,A)}$ functions in (\ref{eq:FiA}) is
weaker than in the case of $\efft=\effb=P=0$.
On the other hand, the errors on the other coefficients are not further
reduced relative to the former situation.

We mention here the effect of the correlation between the constraints from
$\tau_R^{}$ production and those from $\tau_L^{}$ production.
In actual experiments, $\tau_L$ and $\tau_R$ leptons can only be identified on
a statistical basis.
The analyzing power of the semileptonic tau decays is, in principle, equal for
all the semileptonic tau decay modes \cite{Kuhn:1995nn}.
By using the $\tau^- \to \nu_\tau \pi^-$ decay mode, we evaluate the effect of
the $\tau_L^{}$-$\tau_R^{}$ correlation on the errors on $c_i^{(V,A)}$.
We find that the errors on $c_i^{(V,A)}$ may be increased by about 20\% in
actual experiments.

\boldmath
\section{Constraints on general $\h ZV$ couplings}
\unboldmath
\cleqn
\label{sec:numerical}

We are now ready to study the sensitivities to the seven general $\h ZV$
coupling constants.
The errors on these couplings are obtained from those on $c_i^{(V,A)}$ by
using (\ref{eq:ciVA}), (\ref{eq:formfactor}) and (\ref{eq:amp}).
We quantitatively analyze the usefulness of electron beam polarization and of
an additional experiment with another beam energy.
For consistency of the analysis that includes the operators through mass
dimension five, we only keep in $c_i^{(V,A)}$ terms linear in the coupling.

\subsection{Real part}

We first discuss the constraints on the real parts of the general $\h ZV$
couplings.
The constraints on the real parts of the $CP$-even couplings are obtained from
the $CP$-even and $CP\tilde{T}$-even coefficients 
$c_1^{(V,A)},\ldots,c_4^{(V,A)}$, while those on the real parts of the
$CP$-odd couplings are obtained from the $CP$-odd and $CP\tilde{T}$-even
coefficients $c_5^{(V,A)}$ and $c_6^{(V,A)}$.

We first present the results for $\sqrt{s}=250$~GeV\@.
The optimal errors on the $\h ZV$ couplings are summarized in
Table~\ref{table:250}.
\begin{table}[phtb]
 \begin{center}
  \caption{Optimal errors on the real parts of the general $\h ZV$ couplings
 at $\sqrt{s} = 250$~GeV\@.}
  \label{table:250}
  \begin{tabular}[tb]{c|lllll}
   \hdick \\[-2.0ex]
   $\efft$ & ~~--- & ~~0.4 & ~~---  & ~~---  & ~~0.4 \\
   $\effb$ & ~~--- & ~~---  & ~~0.2 & ~~---  & ~~0.2 \\
   $|P|$     & ~~--- & ~~---  & ~~---  & ~~0.9 & ~~0.9 \\
   \hline
 $\Re(b_Z + .059a_Z)$ & .0061 & .0036 & .0033 & .0030 & .0029 \\
 $\Re(c_Z + .059a_Z)$ & .013 & .0076 & .0070 & .0061 & .0061 \\
 $\Re b_\gamma$      & .19 & .072 & .053 & .0085 & .0084 \\
 $\Re c_\gamma$      & .12 & .047 & .035 & .0053 & .0052 \\
 $\Re\tilde{b}_Z$   & .012 & .011 & .010 & .010 & .0091 \\
 $\Re\tilde{b}_\gamma$  & .094 & .036 & .026 & .016 & .013 \\
   \hline
  \end{tabular}
 \end{center}
\end{table}
We only gain sensitivity to six combinations of couplings.
As long as we consider experiments at a fixed collider energy,
one combination of couplings cannot be measured.
The unmeasurable combination of couplings is determined from 
(\ref{eq:formfactor}) and reads
\begin{equation}
 a_Z - ( b_Z + c_Z )\frac{\mmz}{2(s+\mmz)}.
  \label{eq:insensitive}
\end{equation}
It is independent of the final-state fermion flavour $f$ and the electron beam
polarization $P$.
We are thus insensitive to this combination for all $Z$-bosons decay modes.
Since $a_Z$ is the dominant part of (\ref{eq:insensitive}), we fix $a_Z$ to
obtain the optimal sensitivities to the remaining six coupling constants
$b_Z$, $c_Z$, $b_\gamma$, $c_\gamma$, $\tilde{b}_Z$ and $\tilde{b}_\gamma$.
The combinations \mbox{$\Re(b_Z +.059a_Z)$} and \mbox{$\Re(c_Z +.059a_Z)$},
which appear in Table~\ref{table:250}, are orthogonal to the unmeasurable
combination (\ref{eq:insensitive}).

For $\efft=\effb=P=0$, we have good sensitivities only to the three $\h ZZ$
couplings $b_Z$, $c_Z$ and $\tilde{b}_Z$, but not to the $\h Z\gamma$
couplings $b_\gamma$, $c_\gamma$ and $\tilde{b}_\gamma$, which is evident 
from Table~\ref{table:250}.
The functions $F_i^{(A)}$ with $i=1,\ldots,6$ are suppressed in magnitude by
the smallness of $A_f$ and the vanishing of $P$, while the functions
$F_i^{(V)}$ with $i=1,\ldots,6$ have unsuppressed parts.

By using any of the three additional techniques, we gain better sensitivities
to the $\h Z\gamma$ couplings because $F_i^{(A)}$ with $i=1,\ldots,6$ are then
less suppressed.
The measurement of the tau helicity with 40\% efficiency reduces the errors on
$b_\gamma$, $c_\gamma$ and $\tilde{b}_\gamma$ by a factor of about 2/5
relative to the case without tau helicity measurement.
Bottom charge identification with 20\% efficiency reduces the errors on these
couplings by a factor of 2/7.
We observe that the tau helicity measurement leads to an improvement
comparable to that for the bottom charge identification.
This may be understood qualitatively from the relation
\begin{equation}
 \sqrt{ \frac{\effb}{\efft}\,
  \frac{\mbox{Br}(Z\to b\bar{b})}{\mbox{Br}(Z\to \tau^-\tau^+)} |A_b|}
  \approx 1.5
  \approx \frac{2/5}{2/7}.
\end{equation}
The electron beam polarization is the most efficient technique for improving
the sensitivities.
It reduces the errors on the $CP$-even ($CP$-odd) $\h Z\gamma$ couplings by a
factor of about 1/20 (1/6).
A qualitative understanding hereof is obtained from the relation
\begin{equation}
 \sqrt{
  \frac{1}{\efft}\, \frac{1}{\mbox{Br}(Z\to \tau^-\tau^+)} |P|}
  \approx 8.2
  \approx \frac{2/5}{1/20},
\end{equation}
for the $CP$-even couplings.
We find from Table~\ref{table:250} that the errors on the $CP$-even $\h ZZ$
couplings are reduced by a factor of 1/2 with these three additional
techniques, while the $CP$-odd $\h ZZ$ couplings are almost unchanged.

For $\efft=\effb=P=0$, the errors on the real parts of the couplings and the
corresponding correlation matrix are
\begin{equation}
 \label{eq:E250P0b0t0}
 \begin{array}[tb]{ccl}
 \Re(b_Z + .059a_Z) &=& 0 \pm  .0061 \\
 \Re(c_Z + .059a_Z) &=& 0 \pm  .013 \\
 \Re b_\gamma      &=& 0 \pm  .19   \\
 \Re c_\gamma      &=& 0 \pm  .12  \\
 \Re \tilde{b}_Z   &=& 0 \pm  .012 \\
 \Re \tilde{b}_\gamma  &=& 0  \pm  .094
 \end{array},
 \qquad
 \left(
  \begin{array}[tb]{rrrrrr}
   \one \\
   -.95& \one \\
   -.86&  .88&  \one \\
    .83& -.89&  -.99&  \one \\
   \zero&  \zero&  \zero&  \zero&  \one \\
   \zero&  \zero&  \zero&  \zero& -.46&  \one
  \end{array}
  \right) .
\end{equation}
There are strong correlations among the errors on $b_Z$, $c_Z$, $b_\gamma$
and $c_\gamma$.
Thus, a certain combination of parameters is more strongly constrained than
the individual parameters.
The eigenvector with the smallest eigenvalue and its error read
\begin{equation}
 \label{eq:constraint.E250P0b0t0}
  \Re( .08a_Z + .90b_Z + .42c_Z  +  .05b_\gamma + .08c_\gamma)
  = 0 \pm .00076 .
\end{equation}
The above combination may be understood qualitatively by observing that an
experiment at $\sqrt{s}=250$~GeV operates near the threshold of the $\h Z$
production process, where $\beta_{\h Z}\approx0$.
Near the threshold, the form factors $h_1^Z$ and $h_1^\gamma$ play a dominant
role in the helicity amplitudes, while the residual form factors are
suppressed by the smallness of $\beta_{\h Z}$.
Specifically, we have
\begin{eqnarray}
 \label{eq:threshold-amp}
  \hat{M}^{\lambda}_\hi
  =
  g_Z \sqrt{2s} \mz \left[ - g_Z \ges D_Z(s) h_1^Z
	    +  e D_\gamma(s) h_1^\gamma \right]
   + O(\beta).
\end{eqnarray}
Furthermore, the sensitivities to the $\h Z\gamma$ couplings are diminished
for $\efft=\effb=P=0$ because $\sin^2\theta_W \approx 1/4$.
Thus, near threshold there is good sensitivity to the following combination of
couplings:
\begin{equation}
 h^Z_1 \approx 1+a_Z
  + 4 b_Z \left(1+ \frac{m_\h}{\mz} \right)
  + 2 c_Z \frac{\mmh }{\mmz} . \\
\end{equation}
This is essentially the combination that appears in the constraint
(\ref{eq:constraint.E250P0b0t0}).

In models with multiple Higgs doublets, including the minimal supersymmetric
extension of the SM (MSSM), the coupling $a_Z$ is modified at the tree level,
while the couplings $b_V$, $c_V$ and $\tilde{b}_V$ only receive corrections at
the loop level.
Thus, we discuss here the sensitivity to $a_Z$ when $b_V=c_V=\tilde{b}_V=0$.
 From (\ref{eq:E250P0b0t0}), we obtain at $\sqrt{s}=250$~GeV
\begin{equation}
 \label{eq:az}
 a_Z = 0 \pm 0.010
\end{equation}
if $b_V=c_V=0$.

For $\efft=40\%$, $\effb=20\%$ and $|P|=90\%$, the errors and correlation
matrix are found to be
\begin{equation}
 \label{eq:E250P9b2t4}
 \begin{array}[tb]{ccl}
 \Re(b_Z + .059a_Z) &=& 0 \pm  .0029 \\
 \Re(c_Z + .059a_Z) &=& 0 \pm  .0061 \\
 \Re b_\gamma      &=& 0 \pm  .0084  \\
 \Re c_\gamma      &=& 0 \pm  .0052  \\
 \Re \tilde{b}_Z   &=& 0 \pm  .0091 \\
 \Re \tilde{b}_\gamma  &=& 0 \pm  .013
 \end{array},
 \qquad
 \left(
  \begin{array}[tb]{rrrrrr}
   \one \\
   -.96& \one \\
   -.08&  .08&  \one \\
   .08& -.09& -.99&  \one \\
   \zero&  \zero&  \zero&  \zero&  \one \\
   \zero&  \zero&  \zero&  \zero& -.09&  \one
  \end{array}
  \right) .
\end{equation}
We obtain similar correlation matrices for the other situations when only one
of the three additional measurements is employed.
The correlations between the $\h ZZ$ and $\h Z\gamma$ couplings then
disappear.
There are still strong correlations between $b_Z$ and $c_Z$ and between
$b_\gamma$ and $c_\gamma$.
The eigenvectors of the two smallest eigenvalues and their errors are
\bmeq
 \label{eq:constraint.E250P9b2t4}
  \Re(.025a_Z +.28b_Z +.14c_Z  +.50b_\gamma  +.81c_\gamma)
  &=& 0 \pm .00068 ,  \label{eq:constraint.E250P9b2t4gamma} \\
  \Re(.074a_Z + .86b_Z +.40c_Z  -.17b_\gamma  -.27c_\gamma)
  &=& 0 \pm .00078 .
\emeq
As in (\ref{eq:threshold-amp}), the form factors $h_1^Z$ and $h_1^\gamma$
become dominant near the threshold, and there is good sensitivity to the 
following two combinations of couplings:
\bmeq
 \label{eq:threshold-formfactors}
 h^Z_1 &\approx& 1+a_Z
  + 4 b_Z \left(1+ \frac{m_\h}{\mz} \right)
  + 2 c_Z \frac{\mmh }{\mmz} , \\
 h^\gamma_1 &\approx& 2 b_\gamma \left(1+ \frac{m_\h}{\mz} \right)
  + 2 c_\gamma \left( \frac{m_\h}{\mz} + \frac{\mmh }{\mmz} \right) .
\emeq
The most-strongly constrained combinations listed in
(\ref{eq:constraint.E250P9b2t4})
are essentially linear superpositions of $h_1^Z-1$ and $h_1^\gamma$ as given
in (\ref{eq:threshold-formfactors}).

We now discuss the sensitivity to $a_Z$ when $b_V=c_V=\tilde{b}_V=0$.
The six-parameter constraints (\ref{eq:E250P9b2t4}) then lead to the
one-parameter constraint
\begin{equation}
 a_Z = 0\pm0.010
\end{equation}
at $\sqrt{s}=250$~GeV\@.
This constraint is same as in (\ref{eq:az}).
The error on $a_Z$ is not diminished by any of the three experimental options.

Figure~\ref{fig:25gam} displays the contours of $\chi^2=1$ (39\% CL) in the
$(b_\gamma,c_\gamma)$ plane for the different modes of experiment.
The other five degrees of couplings have been integrated out.
We observe that there is a strong correlation between $b_\gamma$ and
$c_\gamma$ for all experimental methods.
As mentioned above, the specific combination of $b_\gamma$ and $c_\gamma$ 
contained in (\ref{eq:constraint.E250P9b2t4}) is thus tightly restricted.
We can see from Figure~\ref{fig:25gam} that the individual sensitivities to
$b_\gamma$ and $c_\gamma$ are drastically improved by the electron beam
polarization.
This means that we can obtain strict constrains on any model that predicts
large $HZ\gamma$ couplings by using data from experiments with polarized
electron beams.

Figure~\ref{fig:25z} displays the contours of $\chi^2=1$ in the
$(b_Z^{},c_Z^{})$ plane.
We see that the $HZZ$ couplings are well constrained even if
$\eps_\tau=\eps_b=P=0$.
The three charge and polarization measurements lead to moderate reductions of
the errors on $b_Z^{}$ and $c_Z^{}$.

Figure~\ref{fig:25bctil} shows the contours of $\chi^2=1$ in the
$(\tilde{b}_Z, \tilde{b}_\gamma)$ plane.
The three charge and polarization measurements mainly reduce the error on
$\tilde{b}_\gamma$.
The reduction of the error on $\tilde{b}_\gamma$ is transferred to that on
$\tilde{b}_Z$ via the correlation between $\tilde{b}_Z$ and
$\tilde{b}_\gamma$.

Next, we consider the case of $\sqrt{s}=500$~GeV\@.
The results for $\efft = 40\%$, $\effb = 20\%$ and $P = 90\%$ are
\begin{equation}
 \label{eq:E500P9b2t4}
 \begin{array}[tb]{ccl}
  \Re(b_Z + .016 a_Z) &=&  0 \pm .0015 \\
  \Re(c_Z + .016 a_Z) &=& 0 \pm  .0007 \\
  \Re b_\gamma &=& 0  \pm  .0024 \\
  \Re c_\gamma &=& 0  \pm  .0005 \\
  \Re\tilde{b}_Z &=& 0  \pm  .0042 \\
  \Re\tilde{b}_\gamma &=& 0  \pm  .0052
 \end{array},
 \qquad
 \left(
  \begin{array}[tb]{rrrrrr}
   \one \\
   -.77  & \one \\
   -.09 &  .07  & \one \\
    .07 & -.09 & -.84  & \one \\
    \zero&  \zero&  \zero &  \zero & \one \\
    \zero&  \zero&  \zero &  \zero & -.09  & \one
  \end{array}
  \right) .
\end{equation}
Although the cross section at $\sqrt{s}=500$~GeV is smaller the one at 
$\sqrt{s}=250$~GeV, the errors are reduced because $b_V^{}$, $c_V^{}$ and
$\tilde{b}_V$ are accompanied by a factor of $s/m_Z^2$.
Increasing the CM energy from $\sqrt{s}=250$~GeV to $\sqrt{s}=500$~GeV
reduces the errors on $b_Z$ and $b_\gamma$ by a factor of 1/2 to 1/3 and those
on $c_Z$ and $c_\gamma$ by a factor of 1/7 to 1/10.
This also reduces the errors on the $CP$-odd couplings $\tilde b_Z$ and
$\tilde b_\gamma$ by a factor of 1/2.
The strong correlations are lost because the experiment at $\sqrt{s}=500$~GeV
is far above the threshold, and the various form factors in the helicity
amplitudes are non-negligible.
At $\sqrt{s}=500$~GeV, the sensitivity to $a_Z$ becomes
\begin{equation}
 a_Z = 0 \pm 0.021
\end{equation}
if $b_V$, $c_V$ and $\tilde{b}_V$ are fixed to zero.
This error is larger than the one at $\sqrt{s}=250$~GeV because $a_Z^{}$ is
the coefficient of the renormalizable dimension-four operator.
We conclude that $a_Z$ may be well measured at the CM energy where the cross
section of $\h Z$ production has its maximum.

Finally, we present optimal constraints on the seven parameters in
(\ref{eq:lag}) by combining the analyses at $\sqrt{s}=250$~GeV and
$\sqrt{s}=500$~GeV with $\efft = 40\%$, $\effb = 20\%$ and $P = 90\%$.
Here, we face the problem that our results depend on the integrated
luminosities at the two energies.
At fixed energy, our constraints scale as $L^{-1/2}$.
We can arbitrarily scale our results by changing the nominal value of $L$,
which could include more realistic experimental efficiencies.
Once we combine the analyses at the two energies, our results will depend on
the ratio of the two respective values of $L$, which we cannot fix a priori.
For simplicity, we assume the same luminosity, $L=10$~fb$^{-1}$, at both
energies.
The results are then
\begin{equation}
 \label{eq:250+500}
 \begin{array}[tb]{ccl}
  \Re(b_Z + .066 a_Z)  &=& 0 \pm .0009 \\
  \Re c_Z       &=& 0 \pm .0006 \\
  \Re b_\gamma   &=& 0 \pm .0015 \\
  \Re c_\gamma   &=& 0 \pm .0004 \\
  \Re\tilde{b}_Z       &=& 0 \pm .0038 \\
  \Re\tilde{b}_\gamma      &=& 0 \pm .0049
 \end{array},
 \qquad
 \left(
  \begin{array}[tb]{rrrrrr}
   \one \\
   -.68  & \one \\
   -.08  &  .07 & \one \\
    .06  & -.08 & -.79 & \one \\
    \zero&  \zero&  \zero &  \zero & \one \\
    \zero&  \zero&  \zero &  \zero & -.09 & \one
  \end{array}
  \right) .
\end{equation}
The minimum $\chi^2$ is found to be
\begin{equation}
\label{eq:chi2min}
  \chi^2_{\mathrm{min}} = \left(\frac{a_Z}{0.024}\right)^2 .
\end{equation}
So far, we have considered $a_Z$ as a fixed parameter.
Now, $\chi^2_{\mathrm{min}}$ is a function of $a_Z$, so that we can obtain
optimal constraints on all seven $\h ZV$ couplings.
The result is
\begin{equation}
 \begin{array}[tb]{ccl}
  \Re a_Z &=& 0 \pm .024 \\
  \Re b_Z &=& 0 \pm .0018 \\
  \Re c_Z &=& 0 \pm .0006 \\
  \Re b_\gamma &=& 0 \pm .0015 \\
  \Re c_\gamma &=& 0 \pm .0004 \\
  \Re\tilde{b}_Z &=& 0 \pm .0038 \\
  \Re\tilde{b}_\gamma &=& 0 \pm .0049
 \end{array},
 \qquad
 \left(
  \begin{array}[tb]{rrrrrrr}
   \one \\
   -.87  & \one \\
   -.02 & -.31  & \one \\
   .00  &-.04 &  .07  & \one \\
   .00  & .03 & -.08 & -.79  & \one \\
   \zero & \zero& \zero & \zero & \zero & \one \\
   \zero & \zero&  \zero&  \zero&  \zero& -.09  & \one
  \end{array}
  \right) .
\end{equation}
Now, $a_Z$ is weakly constrained.
There is a strong correlation between $a_Z$ and $b_Z$.
This reflects the fact that the combination $\Re(b_Z + .066 a_Z)$ in
(\ref{eq:250+500}) is more strongly constrained than $\Re b_Z$.

Figure~\ref{fig:abc} illustrates how $a_Z^{}$ is constrained.
As mentioned above, there is a combination of couplings, namely the one in
(\ref{eq:insensitive}), that is not constrained by an experiment at a single
CM energy.
The projections of the cylinder defined by $\chi^2=1$ onto the $(a_Z, b_Z)$,
$(b_Z, c_Z)$ and $(c_Z, a_Z)$ planes are indicated as the stripes between the
dashed (thin solid) lines for $\sqrt{s}=250$ (500)~GeV\@.
Because the direction of the cylinder varies with $\sqrt{s}$, the measurements
at the two energies, $\sqrt{s}=250$~GeV and 500~GeV, lead to individual
constraints on $a_Z$, $b_Z$ and $c_Z$.

So far, we have assumed that $a_Z$ is constant.
In general, $a_Z$ may have some energy dependence,
\begin{equation}
a_Z(s)= a_Z(0) + s a_Z'(0) + O(s^2) .
\end{equation}
The $O(s^2)$ term is neglected in our approximation.
In terms of operators, the derivative term corresponds to a dimension-five
operator and should be exhausted by the effective Lagrangian (\ref{eq:lag}).
In fact, it may be written as a linear combination of the $b_Z$ and $c_Z$
terms because of the operator identity (\ref{eq:Op.Identity}).

\subsection{Imaginary part}

We now discuss the sensitivities to the imaginary parts of the general $\h ZV$
couplings.
Imaginary parts can arise if the couplings are induced by new interactions
involving particles that can be produced at the energy of the considered
experiment.
For consistency, we neglect the absorptive part in the $Z$-boson propagator as
we ignore all absorptive parts in the SM amplitudes.
The constraints on the imaginary parts of the $CP$-even couplings are then
obtained from the $CP$-even and $CP\tilde{T}$-odd coefficient $c_9^{(V,A)}$,
while the constraints on the imaginary parts of the $CP$-odd couplings are
obtained from the $CP$-odd and $CP\tilde{T}$-odd coefficients $c_7^{(V,A)}$
and $c_8^{(V,A)}$.
The results are summarized in Table~\ref{table:250C}.
\begin{table}[phtb]
 \begin{center}
  \caption{Optimal errors on the imaginary parts of the general $\h ZV$
 couplings at $\sqrt{s} = 250$~GeV\@.}
  \label{table:250C}
  \begin{tabular}[tb]{c|lllll}
   \hdick \\[-2.0ex]
   $\efft$ & ~~--- & ~~0.4 & ~~---  & ~~---  & ~~0.4 \\
   $\effb$ & ~~--- & ~~---  & ~~0.2 & ~~---  & ~~0.2 \\
   $|P|$     & ~~--- & ~~---  & ~~---  & ~~0.9 & ~~0.9 \\
   \hline
 $\Im(b_Z-c_Z)$        & .25 & .095 & .071 & .015 & .014 \\
 $\Im(b_\gamma-c_\gamma)$ & .055 & .027 & .023 & .018 & .018 \\
 $\Im\tilde{b}_Z$       & .049 & .018 & .014 & .0026 & .0026 \\
 $\Im\tilde{b}_\gamma$  & .010 & .0050 & .0043 & .0032 & .0032 \\
   \hline
  \end{tabular}
 \end{center}
\end{table}
In contrast to the real parts of the $\h ZV$ couplings, one can only measure
four combinations of their imaginary parts.
The other combinations, $\Im a_Z$, $\Im(b_Z+c_Z)$ and 
$\Im(b_\gamma+c_\gamma)$, only affect the form factor $h_1^V$, but not $h_2^V$
or $h_3^V$.
They contribute to the amplitudes as a common overall phase, and hence they do
not alter the extracted values of $c_i^{(V,A)}$.
Thus, these other combinations cannot be measured.

For $\efft = \effb = P = 0$, the functions $F_i^{(V)}$ with $i=7,8,9$ are
suppressed in magnitude by the smallness of $A_f$ and the vanishing of $P$,
while the functions $F_i^{(A)}$ with $i=7,8,9$ have unsuppressed parts.
Thus, the errors on \mbox{$\Im(b_\gamma-c_\gamma)$} and
$\Im\tilde{b}_\gamma$ are much smaller than those on \mbox{$\Im(b_Z-c_Z)$} and
\mbox{$\Im\tilde{b}_Z$}, respectively, as may be seen in
Table~\ref{table:250C}.

By using any of the three charge and polarization measurements, we gain better
sensitivities to the $\h ZZ$ couplings because the functions $F_i^{(V)}$ with
$i=7,8,9$ are less strongly suppressed.
The measurement of the tau helicity with 40\% efficiency reduces the errors on
the $\h ZZ$ couplings by a factor of about 2/5.
The bottom charge identification with 20\% efficiency leads to a reduction by
a factor of 2/7.
The electron beam polarization is the most efficient technique for improving
the sensitivities.
It reduces these errors by a factor of 1/20.
The errors on the $\h Z\gamma$ couplings are reduced by a factor of 1/2 with
tau helicity measurements or bottom charge identification, and by a factor of
1/3 with beam polarization.

For $\efft=\effb=P=0$, the errors and the correlation matrix are 
\begin{equation}
 \begin{array}[tb]{ccl}
 \Im(b_Z-c_Z)           &=&  0 \pm .25 \\
 \Im(b_\gamma-c_\gamma) &=&  0 \pm .055 \\
 \Im\tilde{b}_Z        &=&  0 \pm .049 \\
 \Im\tilde{b}_\gamma   &=&  0 \pm .010
 \end{array},
 \qquad
 \left(
  \begin{array}[tb]{rrrr}
   \one \\ 
   -.94  & \one \\
    \zero &  \zero & \one \\
    \zero &  \zero & -.95  & \one
  \end{array}
  \right) .
\end{equation}
There are strong correlations between the errors on the first two terms and
those of the latter two.
The eigenvectors of the two smallest eigenvalues and their errors read
\bmeq
  .20~ \Im\tilde{b}_Z + .98~ \Im\tilde{b}_\gamma \qquad \quad
  &=& 0 \pm  .0031 , \\
  .20~ \Im(b_Z-c_Z) + .98~ \Im(b_\gamma-c_\gamma)
  &=& 0 \pm  .018 .
\emeq
For $\efft = 40\%$, $\effb = 20\%$ and $P = 90\%$, we have
\begin{equation}
 \begin{array}[tb]{ccl}
 \Im(b_Z-c_Z)           &=&  0 \pm .014 \\
 \Im(b_\gamma-c_\gamma) &=&  0 \pm .018 \\
 \Im\tilde{b}_Z        &=&  0 \pm .0026 \\
 \Im\tilde{b}_\gamma   &=&  0 \pm .0032
 \end{array},
 \qquad
 \left(
  \begin{array}[tb]{rrrr}
   \one \\ 
   -.10  & \one \\
    \zero &  \zero & \one \\
    \zero &  \zero & -.10  & \one
  \end{array}
  \right) .
\end{equation}
The strong correlation between the $\h ZZ$ and $\h Z\gamma$ couplings is lost
for the three charge and polarization measurements.

Next, we consider the CM energy $\sqrt{s}=500$~GeV\@.
For $\efft = 40\%$, $\effb = 20\%$ and $|P| = 90\%$, we find
\begin{equation}
 \begin{array}[tb]{ccl}
 \Im(b_Z-c_Z)           &=&  0 \pm .0033 \\
 \Im(b_\gamma-c_\gamma) &=&  0 \pm .0037 \\
 \Im\tilde{b}_Z        &=&  0 \pm .0015 \\
 \Im\tilde{b}_\gamma   &=&  0 \pm .0017
 \end{array},
 \qquad
 \left(
  \begin{array}[tb]{rrrr}
   \one \\ 
   -.10  & \one \\
   \zero & \zero & \one \\
   \zero & \zero & -.10  & \one
  \end{array}
  \right) .
\end{equation}
Although the cross section at $\sqrt{s}=500$~GeV is smaller than the one at
$\sqrt{s}=250$~GeV, the errors on the couplings are reduced because of the
$s/m_Z^2$ factors multiplying the above couplings.
The three combinations $\Im a_Z$, $\Im(b_Z+c_Z)$ and $\Im(b_\gamma+c_\gamma)$
cannot be measured even if the CM energy is varied.

\section{Conclusion}
\cleqn
\label{sec:conclusion}

In the present paper, we have performed a systematic study of the angular
distributions of the process $e^+e^- \to \h f\bar{f}$ in order to assess the
sensitivities to the seven general $\h ZV$ couplings by using the
optimal-observable method
\cite{Atwood:1992ka,Davier:1993nw,Diehl:1994br,Gunion:1996vv}.
To that end, we have expanded the differential cross section as a sum of the
products of the eighteen model-dependent coefficients
$c_i^{(V,A)}$, which contain all the dynamical information on the $\h ZV$
couplings, and the corresponding eighteen angular functions $F_i^{(V,A)}$,
which depend on the production and decay kinematics, the final-state fermion
flavor $f$, the tau polarization and the electron beam polarization $P$.

As for the real parts of the $\h ZV$ couplings, one can only measure six
combinations at a given CM energy $\sqrt{s}$.
At $\sqrt{s}=250$~GeV, we gain optimal errors of order $1\times10^{-2}$
($1\times 10^{-1}$) for the $\h ZZ$ ($\h Z\gamma$) couplings assuming
$L=10 ~\fb^{-1}$ and $\mh=120$~GeV\@.
A tau helicity measurement with 40\% efficiency reduces the optimal errors on
the $\h Z\gamma$ couplings to about 2/5 of those obtainable without such a
measurement.
A bottom charge identification with 20\% efficiency reduces these errors to
about 2/7 of those with unidentified bottom charge.
An electron beam polarization of 90\% reduces the optimal errors on the
$CP$-even ($CP$-odd) $\h Z\gamma$ couplings to about 1/20 (1/6) of those with
unpolarized beams.
The reduction of the errors on the $\h ZZ$ couplings is at most by 1/2.
The sensitivities to the $\h ZV$ couplings depend on $\sqrt{s}$.
The errors on the real parts of the $\h ZV$ couplings decrease by a factor of
about 1/2 to 1/10 when one increases $\sqrt{s}$ from 250~GeV to 500~GeV\@.

As for the imaginary parts of the $\h ZV$ couplings, we can only measure four
combinations, as long as we only keep terms linear in the couplings.
Without the three charge and polarization measurements, we achieve optimal
errors of order $1 \times10^{-2}$ for $\Im\tilde{b}_\gamma$,
$5 \times 10^{-2}$ for $\Im(b_\gamma-c_\gamma)$ and $\Im\tilde{b}_Z$, and
$3 \times 10^{-1}$ for $\Im(b_Z-c_Z)$ with $L=10 ~\fb^{-1}$ at
$\sqrt{s}=250$~GeV\@.
The optimal errors on the $\h ZZ$ couplings are reduced by factors of about
2/5, 2/7 and 1/20 with the tau helicity measurement, the bottom charge
identification and the electron beam polarization, respectively.
The errors on the $\h Z\gamma$ couplings are at most diminished by a factor of
1/3 with the three charge and polarization measurements.
When one increases $\sqrt{s}$ from 250~GeV to 500~GeV, the errors decrease by
a factor of 1/2 to 1/5.

In our analysis, we have considered the general $\h ZZ$ and $\h Z\gamma$
interactions.
We have neglected the contribution from the dimension-five $\h Zee$
operator,
\begin{equation}
\label{eq:dim5}
 \frac{1}{m_Z^{}} \sum_{\sigma=\pm}
  g_\sigma^{\h Zee} \h Z^\mu \bar{e} \gamma_\mu P_\sigma e,
\end{equation}
with $P_\pm=(1\pm\gamma_5)/2$, which contributes to the cross section at the
same order as the operators in the effective Lagrangian (\ref{eq:lag}).
The simple and very general treatment of the observables in
Sect.~\ref{sec:optimal} is no longer valid if the terms in (\ref{eq:dim5}) are
significant.
The optimal constraints on the effective couplings can still be obtained by
directly studying the $g_\sigma^{\h Zee}$ dependences of the differential
cross sections.
We believe, however, that our approach will be useful in constraining theories
that affect the $\h ZV$ couplings more significantly than the $\h Zee$
couplings.
The contributions from the third-generation squarks in the MSSM
\cite{Pilaftsis:1999qt} provide one an example.

\paragraph{Acknowledgments:}
S.I. and B.A.K. are grateful to the members of the KEK theory group for their
hospitality.
The work of S.I. was supported in part by the Ministry of Education, Science,
Sports and Culture of Japan through a Grant-in-Aid in the Visiting Program.
The work of K.H. and J.K. was supported in part through a Grant-in-Aid for the
Special Project Research on Physics of $CP$ Violation.
The work of B.A.K. was supported in part by the German Bundesministerium f\"ur
Bildung und Forschung under Contract No.\ 05~HT9GUA~3, and by the European
Commission through the Research Training Network {\it Quantum Chromodynamics
and the Deep Structure of Elementary Particles} under Contract
No.\ ERBFMRXCT980194.

\appendix

\begin{figure}[ht]
 \begin{center}
 \leavevmode \psfig{file=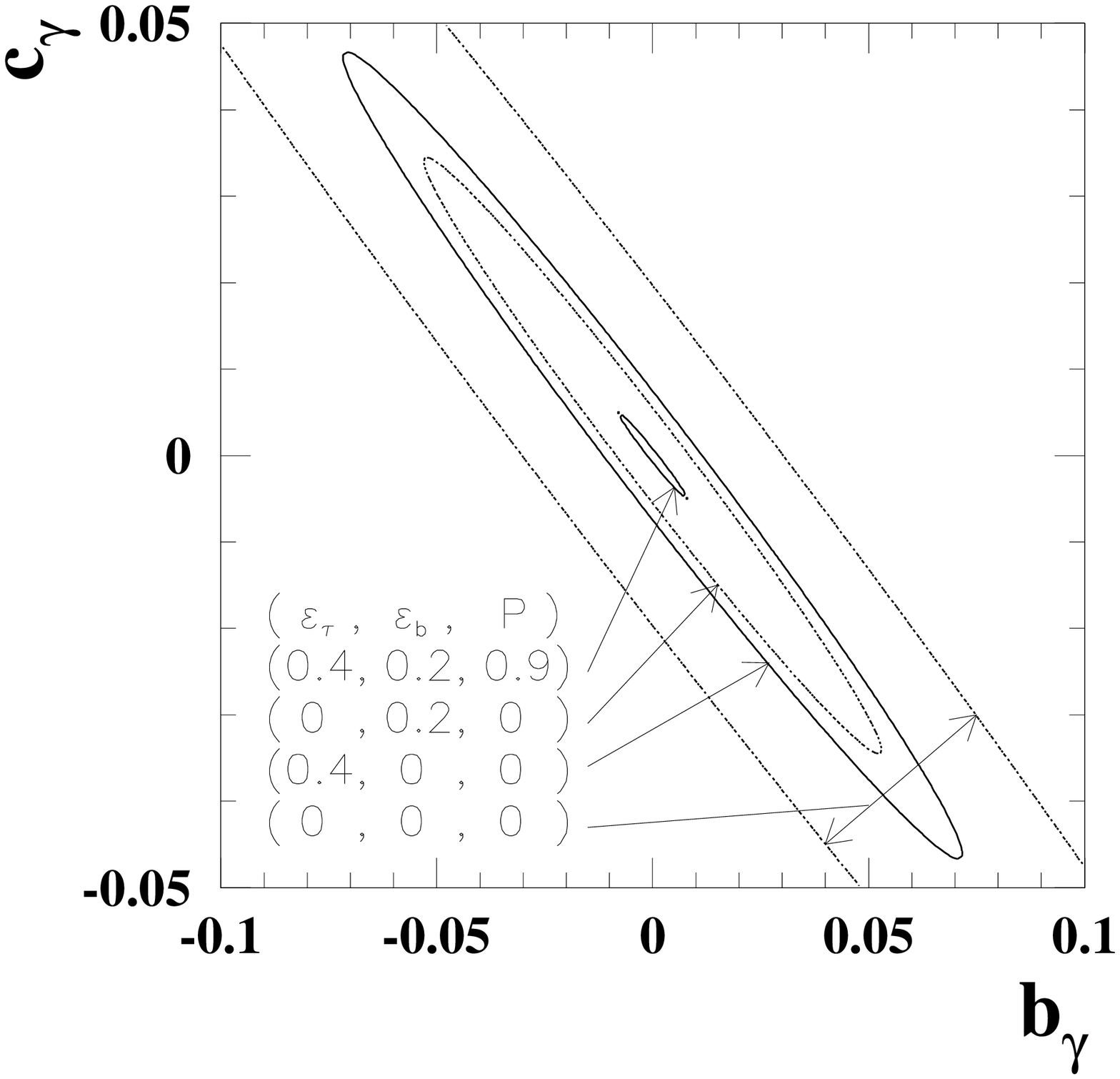,width=16cm}
 \caption[]{}
 \label{fig:25gam}
 \end{center}
\end{figure}

\begin{figure}[ht] 
 \begin{center}
 \leavevmode \psfig{file=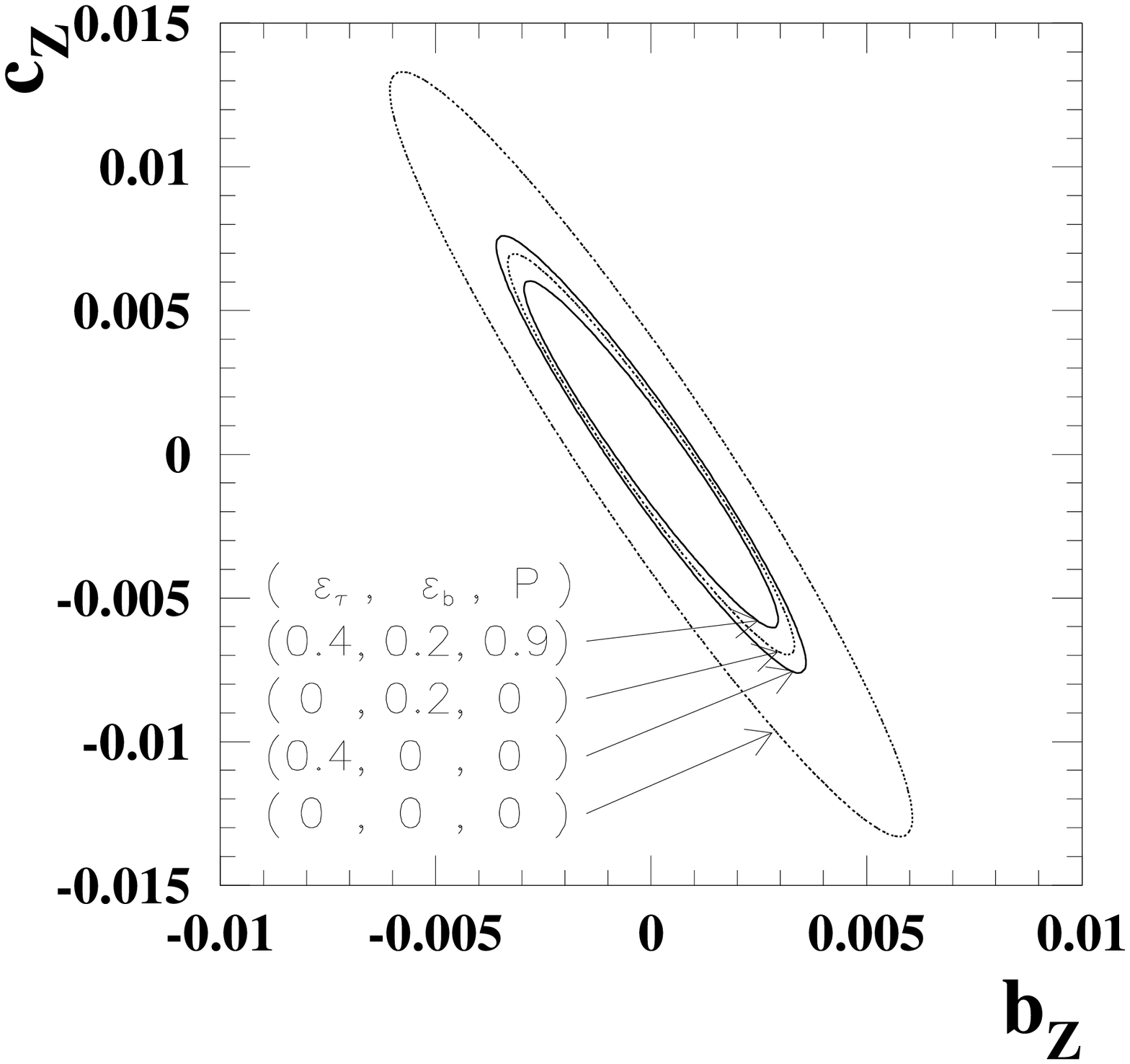,width=16cm}
 \caption[]{}
 \label{fig:25z}
 \end{center}
\end{figure}

\begin{figure}[ht]
 \begin{center}
 \leavevmode \psfig{file=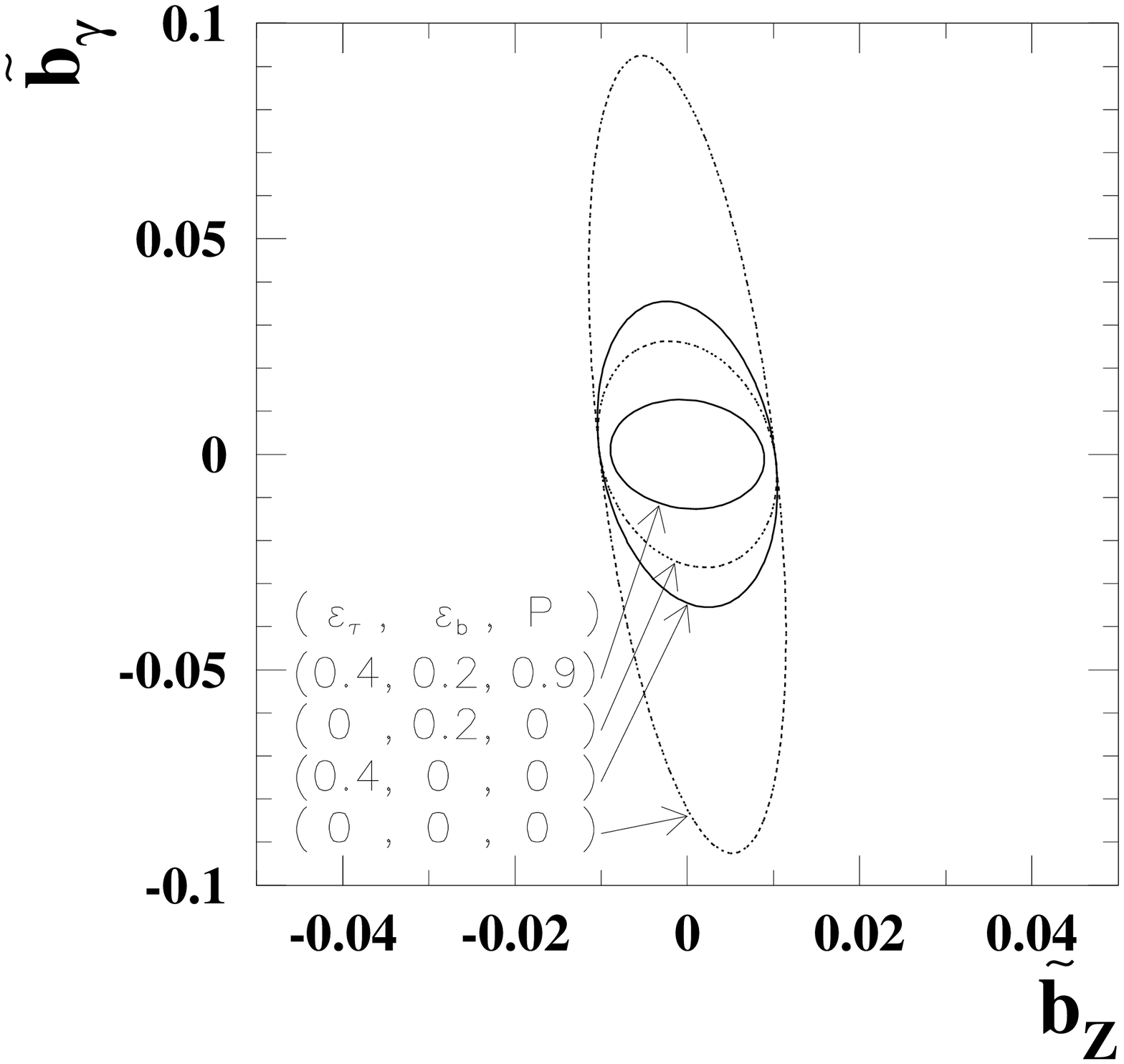,width=16cm}
 \caption[]{}
 \label{fig:25bctil}
 \end{center}
\end{figure}

\begin{figure}[ht]
 \begin{center}
 \leavevmode \psfig{file=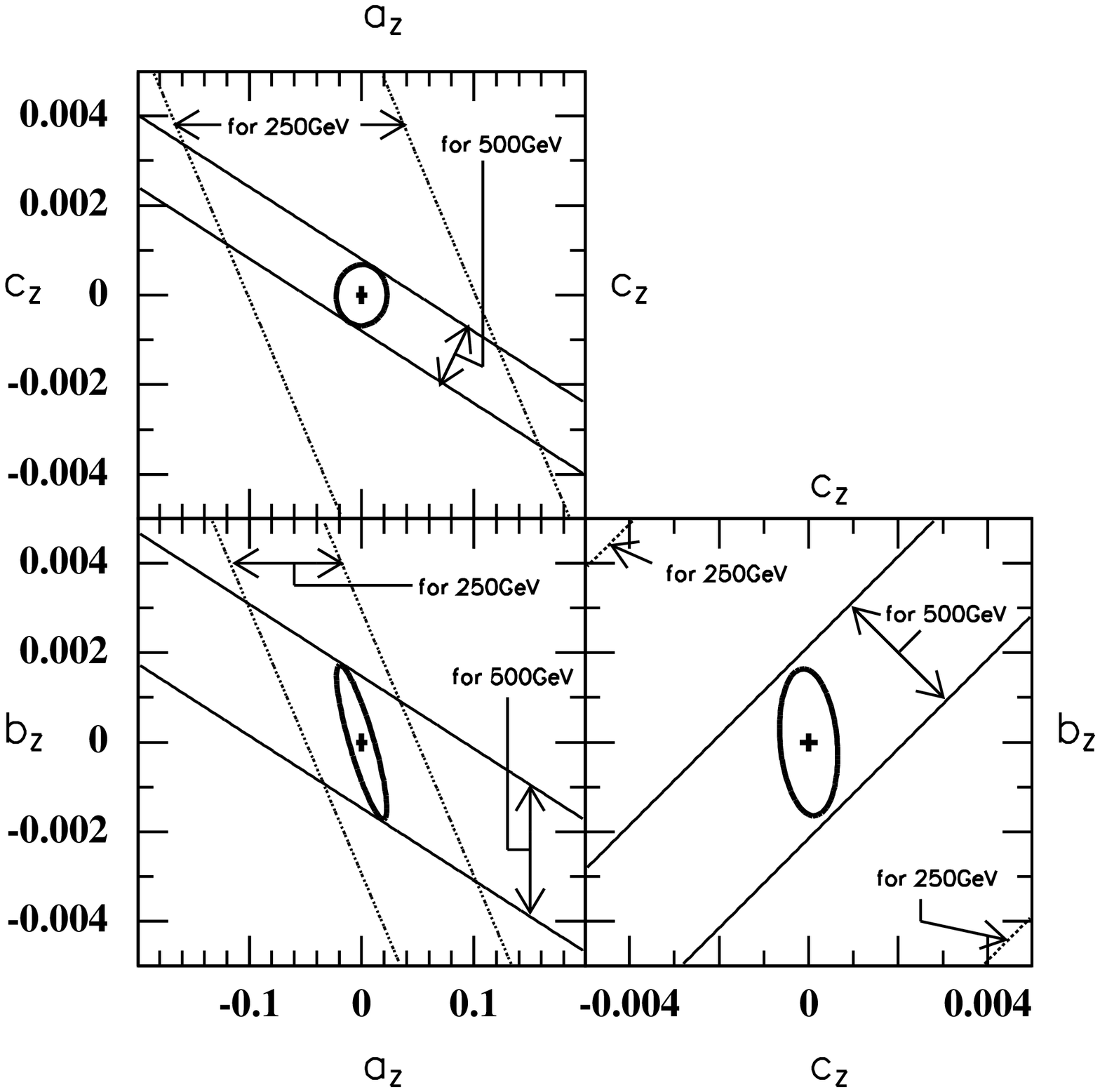,width=16cm}
 \caption[]{}
 \label{fig:abc}
 \end{center}
\end{figure}

\newpage
\begin{center}{\bf Figure captions}
\end{center}
\begin{description}
  \item[Figure~\ref{fign:pflow}:] 
	     {{General $\h ZV$ coupling. 
	     The arrows indicate the direction of the four-momentum flow.} }
  \item[Figure~\ref{fig:25gam}:] 
     Contours of $\chi^2=1$ in the $(b_\gamma, c_\gamma)$ plane for
     $\sqrt{s}=250$~GeV\@.
     The other degrees of general couplings are integrated out.
     The central values of $b_\gamma$ and $c_\gamma$ are assumed to coincide 
     with their SM values, $b_\gamma=c_\gamma=0$.
     The errors are estimated by means of the optimal-observable method 
     under the following four conditions:
     (a) $\efft = \effb = P = 0$;
     (b) $\efft=0.4$ and $\effb=P=0$;
     (c) $\effb=0.2$ and $\efft=P=0$; or
     (d) $\efft=0.4$, $\effb=0.2$ and $|P|=0.9$.
  \item[Figure~\ref{fig:25z}:]
     Contours of $\chi^2=1$ in the $(b_Z, c_Z)$ plane for $\sqrt{s}=250$~GeV\@.
     $a_Z=0$ and the other degrees of general couplings are integrated out.
     The central values of $b_Z$ and $c_Z$ are assumed to coincide with their
     SM values, $b_Z=c_Z=0$.
  \item[Figure~\ref{fig:25bctil}:]
     Contours of $\chi^2=1$ in the $(\tilde{b}_Z, \tilde{b}_\gamma)$
     plane for $\sqrt{s}=250$~GeV\@.
     The other degrees of general couplings are integrated out.
     The central values of $\tilde{b}_Z$ and $\tilde{b}_\gamma$ are assumed to
     coincide with their SM values, $\tilde{b}_Z = \tilde{b}_\gamma = 0$.
  \item[Figure~\ref{fig:abc}:]
     The projections of the $\chi^2=1$ contours
     onto the $(a_Z, b_Z)$, $(b_Z, c_Z)$ and $(c_Z, a_Z)$ planes
     for $\sqrt{s}=250$ GeV and 500 GeV\@. 
     The central values of $a_Z$, $b_Z$ and $c_Z$ are assumed to
     coincide with their SM values, $a_Z=b_Z=c_Z=0$.
     The errors are estimated under the condition
     $\efft=0.4$, $\effb=0.2$ and $|P|=0.9$.
     Dashed (thin solid) lines represent the one-$\sigma$ contours obtained at
     $\sqrt{s}=250$ (500)~GeV\@.
     Thick solid curves represent the combined one-$\sigma$ contours.
\end{description}

\newpage
\begin{center}
 \begin{figure}[ht]
  \begin{center}
  \begin{picture}(350,90)(0,0)
    \Text(-20,50)[]{$V$}
    \Text(0,50)[]{$\beta$}
    \Photon(10,50)(70,50){4}{4}
         \Text(30,70)[]{$\pv$}
         \LongArrow(15,60)(50,60)
    \Vertex(70,50){3} 
    \Photon(70,50)(110,90){4}{5.5}
         \Text(90,100)[]{$p_Z^{}$}
         \LongArrow(80,80)(100,95)
    \DashLine(70,50)(110, 0){4} 
         \Text(90,10)[]{$p_\h^{}$}
         \LongArrow(80,30)(100,5)
    \Text(130,100)[]{$\alpha$}    \Text(150,100)[]{$Z$}
    \Text(130,0)[]{$\h$}
    \Text(150,50)[l]{
   $ i \Gamma^V_{\alpha\beta} =  i \gZ \mz
   \left[
   h^V_1 g_{\alpha\beta}
   +  \frac{h^V_2}{\mmz} \pv_\alpha p_{Z\beta}^{}
   +  \frac{h^V_3}{\mmz} \epsilon_{\alpha\beta\mu\rho} \pv^\mu p_Z^\rho
   \right]
   $ }
  \end{picture}
  \end{center}
\caption[]{}
    \label{fign:pflow}
 \end{figure}
\end{center}

\end{document}